\documentclass[structabstract]{aa}  
\usepackage{graphicx}
\usepackage{natbib}
\bibpunct{(}{)}{;}{a}{}{,} % to follow the A&A style
\usepackage{txfonts}
\begin{document}
   \title{Constraining the nature of the most distant gamma-ray burst host galaxies\thanks{Based on observations made with ESO Telescopes at the Paranal Observatory under programs 60.A-9402(A), 085.A-0418(A), and 085.A-0418(B).}}

  % \subtitle{I. Overviewing the $\kappa$-mechanism}

   \author{
        S. Basa\inst{1},
        J.G. Cuby\inst{1},
        S. Savaglio\inst{3},
        S. Boissier\inst{1},
        B. Cl\'ement\inst{1,4},
        H. Flores\inst{2},
        D. Le Borgne\inst{5}
                \and
       A. Mazure\inst{1}
            }

   \institute{Laboratoire d'Astrophysique de Marseille - LAM, Universit\'e Aix-Marseille \& CNRS, UMR7326, 38 rue F. Joliot-Curie, 13388 Marseille Cedex 13, France\\
                 \email{stephane.basa@oamp.fr}
         \and
                 GEPI-Observatoire de Paris Meudon. 5 Place Jules Jannsen, F-92195, Meudon, France
          \and
                 Max Planck Institute for Extraterrestrial Physics, 85748 Garching bei M\"unchen, Germany
         \and
                Steward Observatory, University of Arizona, 933 N. Cherry Ave, Tucson, AZ 85721, USA
         \and
                 UPMC-CNRS, UMR7095, Institut d'Astrophysique de Paris, 98 bis boulevard Arago, F-75014, Paris, France 
                }
 %\offprints{stephane.basa@oamp.fr}

   \date{Received September 15, 1996; accepted March 16, 1997}

% \abstract{}{}{}{}{} 
% 5 {} token are mandatory
 
  \abstract
   {}
  {Long duration gamma-ray bursts (GRBs) allow us to explore the distant Universe, and are potentially the most effective tracer of the most distant objects. Our current knowledge of the properties of GRB host galaxies at redshifts $\gtrsim5$ is very scarce. We propose to improve this situation by obtaining more observations of high-redshift hosts to better understand their properties and help enable us to use GRBs as probes of the high-redshift universe.}
  {We performed very deep photometric observations of three high-redshift GRB host galaxies, GRB 080913 at $z =6.7$, GRB 060927 at $z =5.5$ and GRB 060522 at $z =5.1$. Our FORS2 and HAWK-I observations at the Very Large Telescope (VLT) targeted the rest-frame ultraviolet continuum of these galaxies, allowing us to constrain their star formation rates (SFRs). In addition, we completed deep spectroscopic observations of the GRB080913 host galaxy with X-Shooter at the VLT to search for Ly-$\alpha$ emission. For the sake of the discussion, we use published results on another high-redshift GRB host, GRB 050904 at $z = 6.3$. The sample of GRB host galaxies studied in this paper consists of four out of the five spectroscopically confirmed GRBs at $z > 5$. }
  {Despite our presented observations being the deepest ever reported of high-redshift GRB host galaxies, we do not detect any of the hosts, neither in photometry nor in spectroscopy in the case of GRB 080913. These observations indicate that the GRB host galaxies seem to evolve with time and to have lower SFRs at $z >5$ than they have at $z\lesssim1$. In addition, the host galaxy of GRB 080913 at $z =6.7$ does not show Ly-$\alpha$ emission.} %somewhat in agreement with the observed trend of decreasing Ly-$\alpha$ emission at redshifts $\gtrsim 6.5$.}
   {While the measured properties of the galaxies in our sample agree with the properties of the general galaxy population at $z>5$, our observations are not sufficiently sensitive to allow us to infer further conclusions on whether this specific population is representative of the general one. The characterization of high-redshift GRB host galaxies is a very challenging endeavor requiring a lot of telescope time, but is necessary to improve our understanding of the high-redshift universe using GRB observations.}


\titlerunning{Constraining the nature of the most distant GRB host galaxies}
\authorrunning{S. Basa and al.}

 \maketitle
%
%______________________________________________________________
%
\section{Introduction}

Searching for high-redshift galaxies is a very active field in observational cosmology. Observing the most distant galaxies is important for improving our understanding of the formation and evolution of these objects, and the reionization of the Universe \citep[for a review, see][]{Robertson 2010}

Galaxies at redshifts above seven and up to about eight are now regularly detected from space and ground based wide-field photometry, either with narrow band filters for Ly-$\alpha$ emitters or broad band filters for Lyman break galaxies \citep{Ouchi 2009,McLure 2010,Bouwens 2010,Castellano 2010,Tilvi 2010, Vanzella 2011, Clement 2011}. Spectroscopic confirmation of a handful of these objects has now been reported \citep{Iye 2006,Stark 2010,Lehnert  2010,Fontana 2010,Vanzella 2011,Pentericci  2011,Schenker 2012,Ono 2012}. These observations are of considerable interest because they allow us to constrain the reionization history of the Universe, and to understand the nature of the sources responsible for this reionization. 

Long gamma-ray bursts (GRBs) offer an alternative method and an attractive shortcut in this difficult quest to detect the first galaxies \citep[for a review, see][]{Gehrels 2009}. Thanks to their tremendous X-ray and gamma-ray prompt emission, GRBs can be detected out to very high-redshifts (higher than eight), e.g. from the {\it Swift} and {\it Fermi} satellites. In addition, the  absorption features of the intergalactic medium (IGM) on the otherwise featureless power-law spectra of the GRB afterglows provide spectroscopic tools for the redshift determination and investigation of the IGM properties \citep{Kawai 2006,Greiner 2009}.

%\citep{Bouwens 2011,Finkelstein 2011}

%Searching for high-redshift galaxies is a very active field in observational cosmology. The most distant galaxies provide a  direct probe of the state of the infant Universe,  for instance the cosmic reionization due to the formation of the first gravitationally collapsed objects. Galaxies up to redshift $\sim8$ are now regularly detected thanks to the combination of the Lyman-break technique with Hubble Space Telescope WFC3 infrared camera \citep{McLure 2010,Bouwens 2010}, wide field ground-based surveys \citep{Ouchi 2009,Castellano 2010} and Ly-$\alpha$ emitter search with low-dispersion spectroscopy \citep{Tilvi 2010, Vanzella 2011, Clement 2011}. These observations provide an important advance by accessing the faint side of the galaxy UV Luminosity Function (LF) \citep{Bouwens 2011,Finkelstein 2011}, showing that these galaxies are likely the sources of the reionization of the universe.

%In the difficult quest to detect the first galaxies, long duration gamma-ray bursts (GRBs) offer a very attractive and alternative shortcut. Being incredibly luminous and having  a nearly perfect featureless power-law emission at the source, spectroscopic observation of high-redshift GRBs can easily provide a redshift measurement from  the detection of the Lyman break, bypassing the reliability on faint emission lines. Their potential as beacons to the distant Universe is now widely recognized  \citep[see for example][]{Gehrels 2009}.

However,  we need to understand how the GRB host galaxy population is representative of the whole galaxy population, in general. There are already some indications of the global properties of the GRB host galaxy population at low redshift, $z \lesssim 2$, which appear to be  generally low luminosity \citep{lefloch 2003}, low metallicity \citep{Vreeswijk 2001,Gorosabel 2005,Starling 2005,Wiersema  2007},  and star-forming galaxies \citep{Christensen 2004,Chen 2009}. Their morphologies are very diverse: irregulars, spheroids, spirals, and mergers \citep{Conselice 2005,Wainwright 2007}. The investigation of the stellar populations of these hosts, using multi-band photometry, suggests that GRB hosts at $z \lesssim 2$ are small star-forming galaxies with likely sub-solar metallicity \citep{Savaglio 2009}.%, and that they do not belong to specific galaxy populations. 

Only a partial glimpse has been obtained of the GRB host galaxy population, at somewhat moderate redshifts, $z \lesssim 2$, and very little about the properties of GRB host galaxies at higher redshift is known. More observations are therefore required to enhance our understanding of both their properties and the implications of using GRBs as probes of the high-redshift universe. 

To achieve this aim, we performed very deep photometric observations of three high-redshift GRB host galaxies and deep spectroscopic observations of one of them. Our targets are the hosts of GRB 080913 at $z = 6.7$ \citep{Greiner 2009}, GRB 060927 at $z = 5.5$ \citep{Ruiz-Velasco 2007}, and GRB 060522 at $z = 5.1$ \citep{Cenko 2006}. We used FORS2 and HAWK-I at the Very Large Telescope (VLT) to observe in the near infrared the rest-frame ultraviolet (UV) continuum of these galaxies and derive, or constrain, their star formation rates (SFRs). On one object, GRB 080913, the most distant in our sample, we performed deep spectroscopic observations with X-Shooter at the VLT to search for Ly-$\alpha$ emission.
 
Our observations and data analysis are discussed in Sect. \ref{sec:obs}. We then discuss in Sect. \ref{sec:disc} the constraints derived from our observations on the GRB host galaxy properties at high redshift, $z>5$.

Throughout this paper, we assume a flat $\Lambda$CDM model with $H_0$=70 km.s$^{-1}$ and $\Omega_M$=0.3 \citep{Komatsu 2011}. All magnitudes are measured on the AB system \citep{Oke 1983}.

%
%______________________________________________________________
%
\section{Observations and detection limits}
\label{sec:obs}
 \subsection{Sample selection}
We selected the most distant spectroscopically confirmed GRBs ($z > 5$) observable from the VLT site, which resulted in a sample of five objects. Two of them were excluded from our sample: the host of GRB 050904 at $z = 6.3$, for which deep Hubble Space Telescope and Spitzer Space Telescope observations exist \citep{Berger 2007}, and the host of GRB 090423, which is the most distant of all spectroscopically confirmed GRBs  \citep{Salvaterra 2009, Tanvir 2009}. We decided that the latter object would have been too faint and difficult to observe, decision that was probably wise in the face of our results. 

The final sample finally targeted the host galaxies of GRB 080913 at $z =6.7$  \citep{Greiner 2009}, GRB 060927 at $z =5.5$ \citep{Ruiz-Velasco 2007}, and GRB 060522 at $z =5.1$ \citep{Cenko 2006}. These three GRBs have properties that are typical of the long GRB population with an isotropic energy release $E_{iso} \approx 7-8\times 10^{52}$ erg. All of these redshifts were determined from the spectroscopic observation of the GRB afterglow emission.

Our final sample consists of three out of the five spectroscopically confirmed GRBs at $z > 5$ known to date.
 
 \subsection{Optical/near-infrared photometry}
\subsubsection{Observations}
\begin{figure}
\begin{center}
  \resizebox{\hsize}{!}{\includegraphics{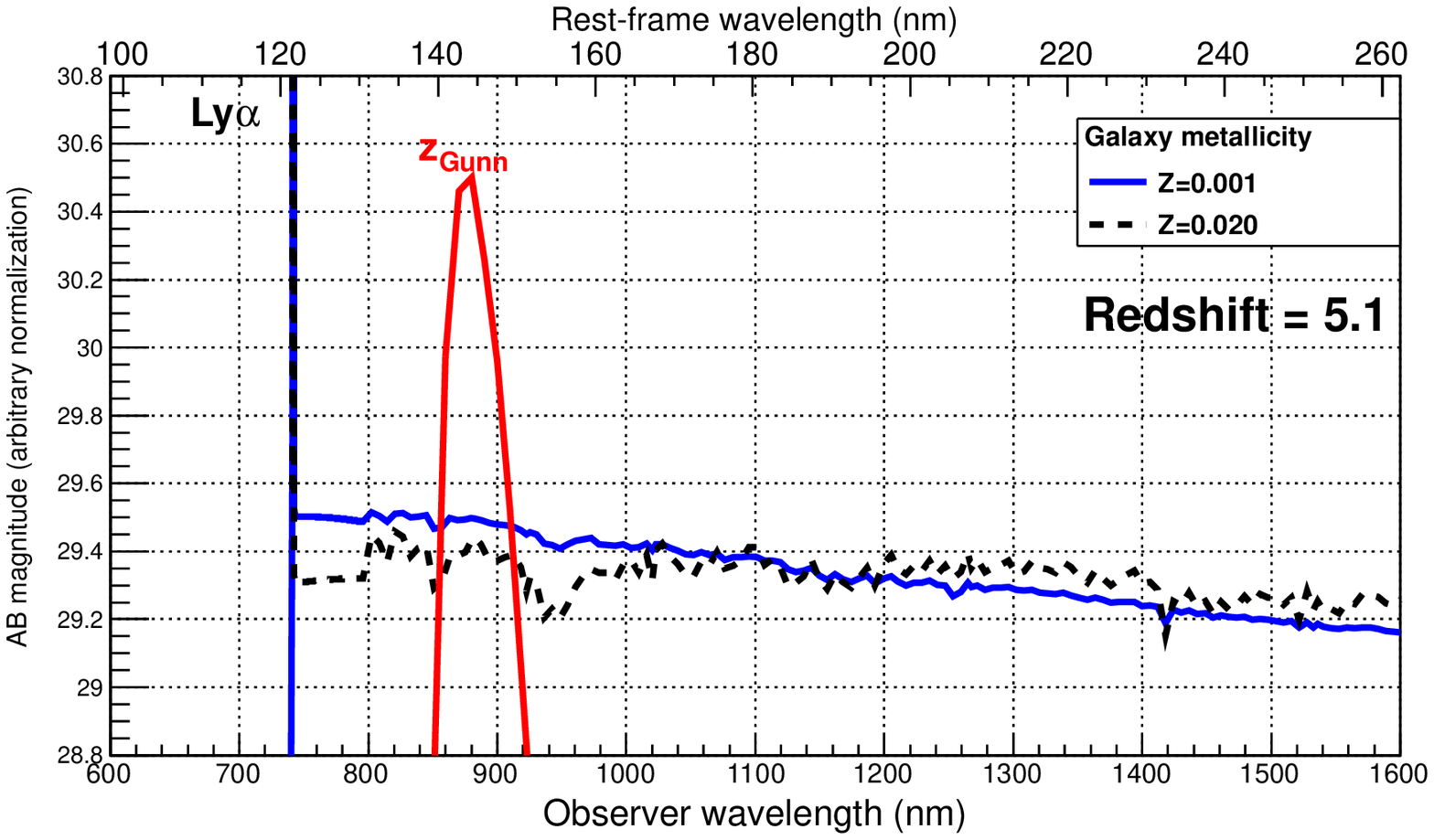}}
  \resizebox{\hsize}{!}{\includegraphics{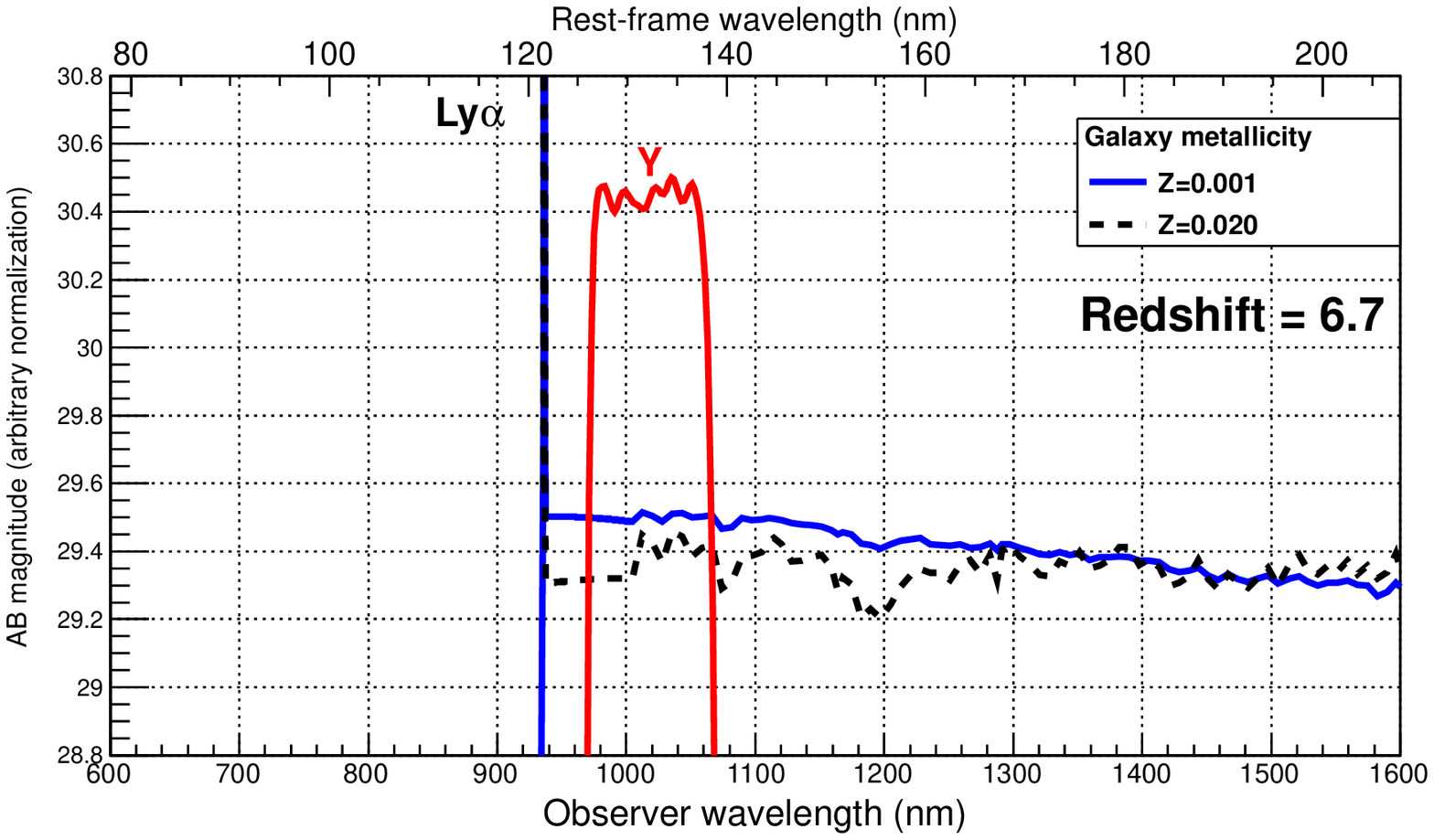}}
\end{center}
\caption{Example of galaxy spectral energy distribution with two different metallicities from \citet{Schaerer 2002}, and with the FORS2 and HAWK-I bands superimposed (magnitudes are arbitrary normalized). To illustrate the band selection, the spectra have been redshifted to 5.1, as for GRB 050622 (top), and 6.7, as for GRB 080913 (bottom).}
 \label{fig:filter}
\end{figure}

The photometric bands were chosen to sample the rest-frame UV continuum of the targets at \hbox{$\sim$1500 {\AA}}, which is longward of the Ly-$\alpha$ forest, but blue enough to minimize any spectral contamination from old stellar populations (figure~\ref{fig:filter}). This domain is well-suited to estimate the SFR. 

We accordingly observed GRB 060927 in the $z_{Gunn}$ band with FORS2, and GRB 060927 and GRB 080913 in the {\it Y} band with HAWK-I. The central wavelengths were $\lambda_0=910$ nm for the $z_{Gunn}$ filter and $\lambda_0=1021$ nm for the {\it Y} band filter. The targets were centered on the detectors providing the optimal performance for their respective instruments, in practice chip\#1 (master CCD) for FORS2 and quadrant\#3 for HAWK-I. Only the data from these detectors were subsequently reduced and analyzed.

The observations were carried out in service mode between July and October 2010. Total exposure times and observing conditions for each target are reported in table \ref{table:0}.

\begin{table*}
\caption{Total exposure times and mean seeing during the photometric observations. Clear to photometric nights were required.}           
\label{table:0}    
\centering
\begin{tabular}{c c c c c c}
\hline\hline
Host galaxy & Redshift & Instrument &Photometric &  T$_{Expo}$ & Mean Seeing    \\   
                    &               &                   & Band           &  (hour)          & (arcsec)            \\   
\hline
GRB 080913 & 6.7 & HAWKI-I & $Y$                    & 8.2 & 0.84''  \\
GRB 060927 & 5.5 & HAWKI-I & $Y$                    & 8.2 & 0.70''  \\
GRB 060522 & 5.1 & FORS2   & $z_{Gunn}$       & 8.9 & 0.83''  \\
\hline
\end{tabular}
\end{table*}

\subsubsection{Data reduction}

The data were reduced using IRAF\footnote{\label{foot:iraf}IRAF is distributed by the National Optical Astronomy Observatory, which is operated by the Association of Universities for Research in Astronomy, Inc., under cooperative agreement with the National Science Foundation.} and AstrOmatic\footnote{See http://www.astromatic.net/} routines, allowing us to control the reduction process step by step. 

One of the most delicate data-reduction steps in the near-infrared domain is sky subtraction. We adopted the prescription described in details in \citet{Clement 2011}, which allows us to estimate the sky at any particular pixel by building a running sky frame for each science frame. A relative astrometric solution is computed for each sky-subtracted frame using Scamp and a fourth-order polynomial fit of bright star positions across the detector plane. All the resulting images are then resampled to a common reference frame using a LANCZOS4 interpolation kernel and finally stacked with Swarp \citep{Bertin 2002}.

\subsubsection{Detection limits}
\label{sec:lim-photo}
\begin{figure*}
\centering
  \includegraphics[width=17cm]{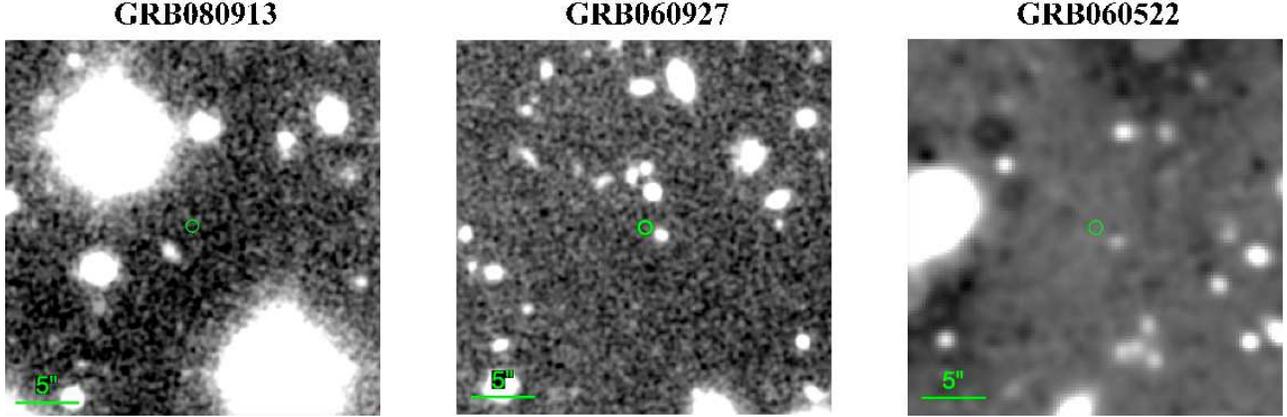}
   \caption{Smoothed image of the fields centered on GRB 080913 (HAWK-I, $Y$ band), GRB 060927 (HAWK-I, $Y$ band), and GRB 060522 (FORS2, $z_{Gunn}$ band). The GRB afterglow localization (circle) based on the optical observations, is relatively accurate (better than 1''). No host galaxy candidate is detected down to the magnitude limits indicated in table \ref{table:1}. Each image covers a field of view of 30''$\times$30''. North is up, east is left.}
   \label{fig:grb}
\end{figure*}

Figure \ref{fig:grb} shows the fields centered on GRB 080913, GRB 060927, and GRB 060522. 

In each of the GRB 060927 and GRB 060522 fields, one object is detected within 2\arcsec\ of the GRB position. At a redshift of $z\sim5$, this corresponds to a distance of $\sim$ 12 kpc, which, even if large, does not allow us to definitively reject these objects as the possible host galaxies of the GRBs. We analyzed archival FORS2 imaging data of these two fields in the $R$ band\footnote{Program: 177.A-0591 (PI: J. Hjorth) and 077.D-0661 (PI: P.M. Vreeswijk).}. The same objects are also detected in this band, which excludes that they are at a redshift higher than $\sim$ 4 and therefore that they are the host galaxies of the GRBs. Therefore, none of the three GRB hosts is detected down to the limiting magnitudes of our observations.

We performed simulations to estimate the limiting magnitudes reached by our observations. Fake point sources with different magnitudes were added to the stacked images. SExtractor  \citep{Bertin 1996} was then used to recover the fake sources, allowing us in turn to estimate the completeness of our observations. The magnitude limits given throughout this paper are the point-source 80\% completeness limits. 

Moreover, in combining the final images for each of the three fields, we tested the impact of the different stacking methods (mean, median, and sigma-clipping) on the detection limits reached by our observations. The limits were identical to within $\sim 0.1$ magnitude, an error that we conservatively added to the photometric errors measured by SExtractor. 

Before converting the measured fluxes to luminosities, we investigate the possible effects of dust, both locally in the Galaxy and in the GRB hosts. The Galactic extinction in the direction of the three fields  is on the order of $<0.1$ magnitude \citep{Schlegel 1998}. As to the dust in the GRB hosts, we assume that the extinction, inferred from the difference between the observed and expected GRB afterglow spectral indices, is representative of the overall extinction in the galaxy. No evidence of dust attenuation was found in the afterglow observation of GRB 060927 and GRB 080913 \citep{Greiner 2009,Ruiz-Velasco 2007}. Short of any estimate of the extinction for GRB 060522, we assume that there is no extinction in this object.

\begin{figure}
  \resizebox{\hsize}{!}{\includegraphics{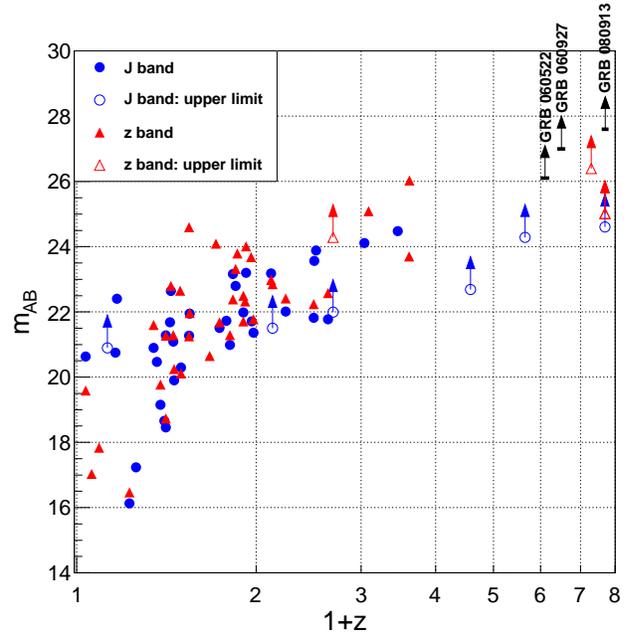}}
  \caption{Magnitude limits for GRB 080913 ($Y$ band), GRB 060927 ($Y$ band), and GRB 060522 ($z_{Gunn}$ band). Magnitudes in the J and $z$ bands of known GRB host galaxies compiled from the GHostS database are also reported.}
  \label{fig:mag}
\end{figure}

The final detection limits of our three host galaxies are reported in table \ref{table:1} and figure~\ref{fig:mag} together with the magnitudes of known GRB hosts at lower redshifts. To illustrate the depth reached by our data, the limiting magnitude on GRB 080913 host data is for instance $\sim$ 1.5 magnitude deeper than previously published observations of this object.

\begin{table*}
\caption{Magnitude limits for the three GRB host galaxies. An estimate of the Galactic extinction in the bands used for the observations is indicated.}           
\label{table:1}    
\centering
\begin{tabular}{c c c c c}
\hline\hline
Host galaxy & Redshift & Band &  Magnitude limit &  Galactic extinction\\   
                    &               &          &   (AB mag) &    (AB mag)\\   
 \hline
GRB 080913 & 6.7 &  $Y$                &  $> 27.6$  & 0.05 \\
GRB 060927 & 5.5 &  $Y$                &  $> 27.0$  & 0.07 \\
GRB 060522 & 5.1 &  $z_{Gunn}$   &  $> 26.1$ & 0.07 \\
\hline
\end{tabular}
\end{table*}

\subsection{Optical/near-infrared spectroscopic observations}
\subsubsection{Observations}
Deep spectroscopic observations of GRB 080913 at $z = 6.7$ were performed during the X-Shooter science verification, with the aim of searching for possible Ly-$\alpha$ emission in the spectrum of the host galaxy. 

We used X-Shooter in the integral field unit (IFU) mode. The three-slice IFU has a total field of view of 4''$\times$1.8'' and the centering was done on the GRB position, allowing between a 1\arcsec\ and 2\arcsec\ uncertainty in the GRB and the host galaxy positions. The IFU field of view was re-imaged at the entrance of the spectrograph as three $0.6 \times 4$\arcsec sub-slits.

Observations  were carried out in service mode in September and October 2009. Twenty exposures of 570 s each were taken under excellent observing conditions (mean seeing $\approx 0.8''$ and clear to photometric nights) for a total integration time of 3.2 h.

\subsubsection{Data reduction}

The first steps of the processing of the IFU VIS arm observations was performed using the X-Shooter data reduction pipeline version 1.2.2 \citep{Goldoni 2006}, namely sky subtraction, distortion correction, and flux calibration. The 20 two-dimensional images were then stacked using standard IRAF routines. Standard star observations were used to complete the spectrophotometric calibration. The reconstruction of the datacube was carried out following the procedure described in \citet{Flores 2011}.

The final spectrum covers the wavelength range [550-1020] nm, with a spectral resolution of R$\sim$13200.

\subsubsection{Detection limit}
\label{sec:lim-spectro}
\begin{figure}
  \resizebox{\hsize}{!}{\includegraphics{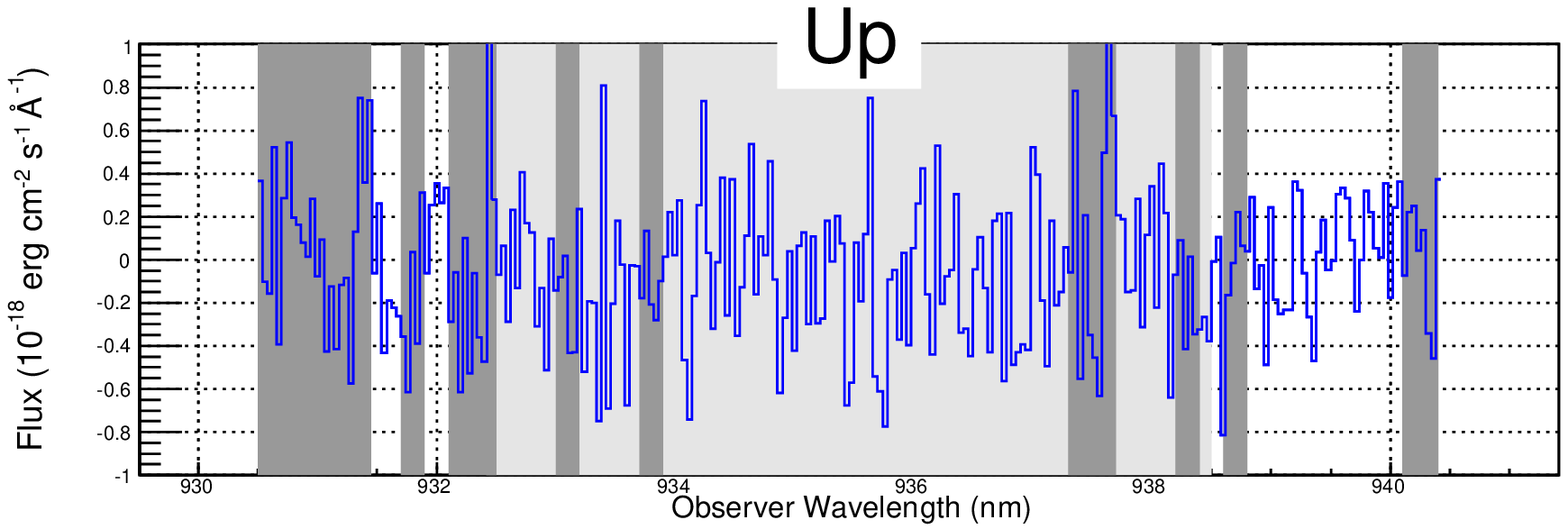}}
  \resizebox{\hsize}{!}{\includegraphics{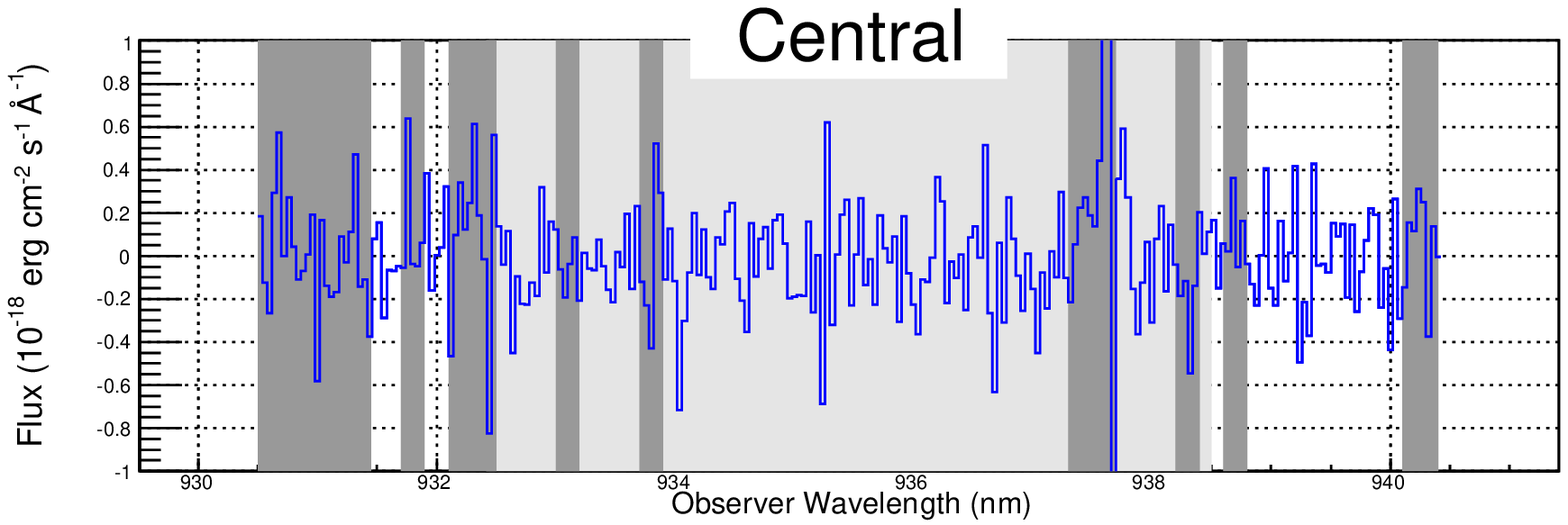}}
  \resizebox{\hsize}{!}{\includegraphics{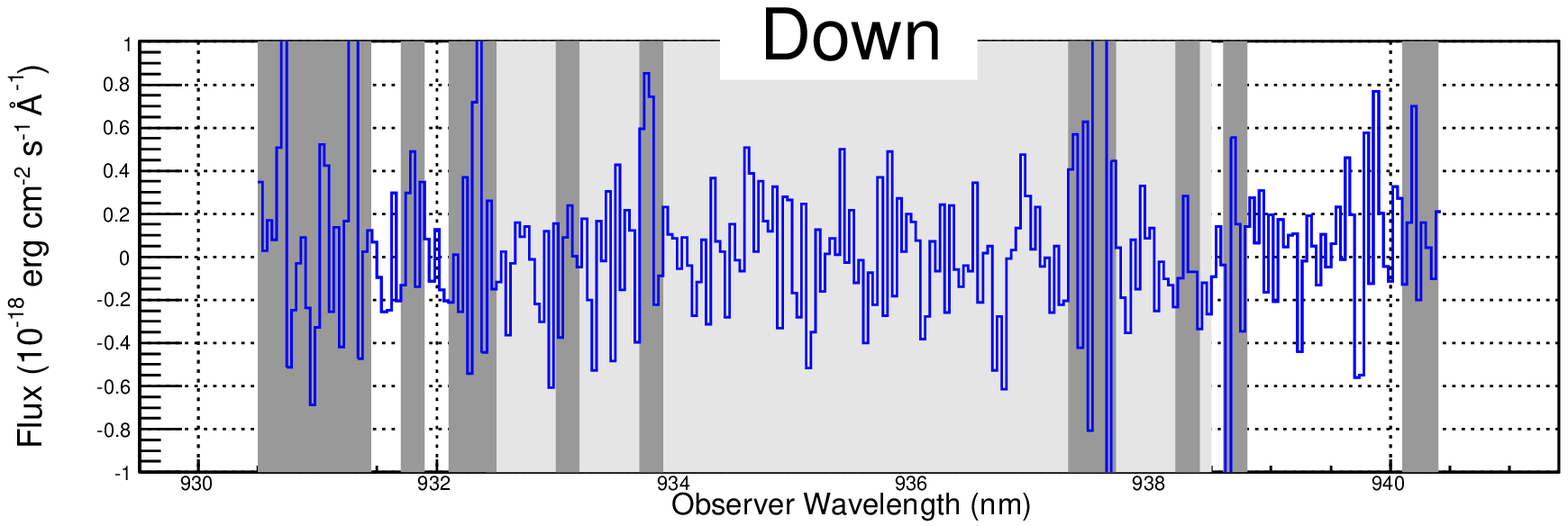}}
  \resizebox{\hsize}{!}{\includegraphics{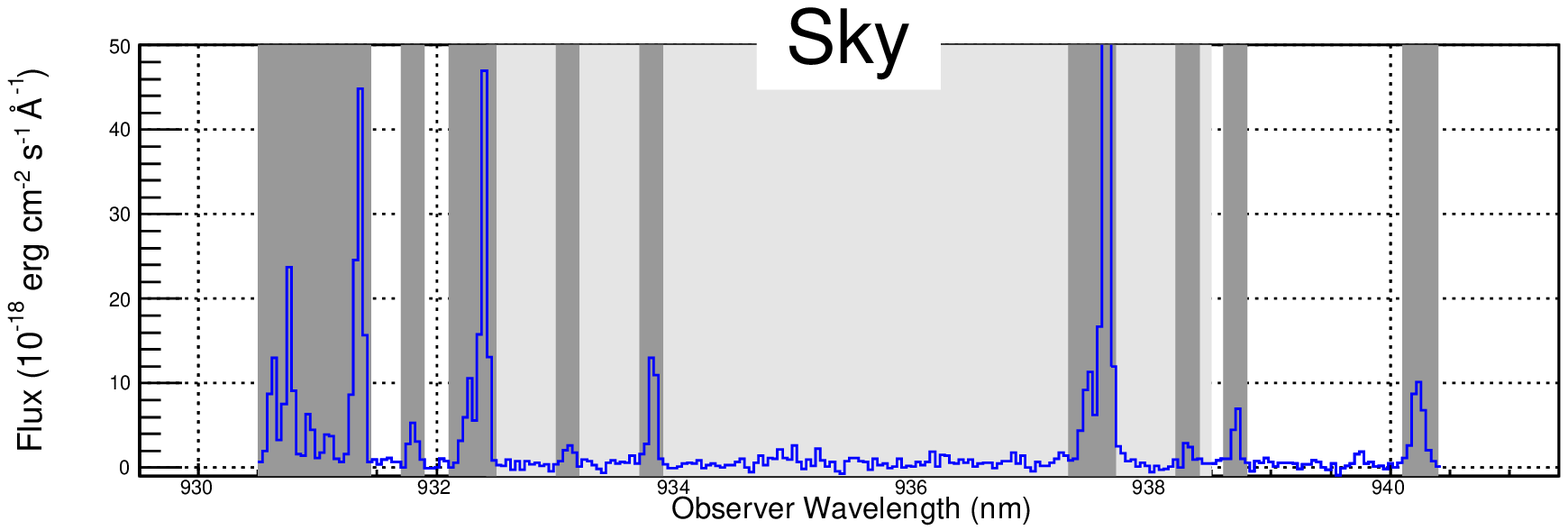}}
 \caption{Sky-subtracted spectra of GRB 080913 at $z =6.7$ for the three sub-slits of the VLT/X-Shooter IFU. The central filled area (light blue) indicates the domain where the Ly-$\alpha$ emission line is expected, within the uncertainty in the redshift of the object \citep{Greiner 2009}. The width of the light blue area is twice the redshift error. The sky spectrum is displayed at the bottom of the figure and the positions of the main sky emission lines are indicated by the gray areas.}
  \label{fig:spec}
\end{figure}

Figure \ref{fig:spec} shows the sky-subtracted spectrum of GRB 080913 at $z =6.7$. Various filtering methods such as weighted Gaussians or wavelets were tested but did not allow us to detect the presence of an emission line in any of the three IFU slices.

The continuum is not detected as expected from the photometric and spectroscopic detection limits reached by our data: M$_{Y}$ $>27.6$ (table \ref{table:1}) corresponds to $\lesssim9 \times $10$^{-21}$ erg~cm$^{-2}$~s$^{-1}$~{\AA}$^{-1}$ at 1021 nm, while the spectroscopic detection limit of our X-Shooter data is $\sim0.8 \times$10$^{-18}$ erg~cm$^{-2}$~s$^{-1}$~{\AA}$^{-1}$.

To estimate the limit imposed by our null detection of a Ly-$\alpha$ emission line, we performed simulations of emission lines based on the properties of z$\approx$6.6 Ly-$\alpha$ emitters (LAEs), at a redshift very similar to the redshift of GRB 080913. The average full width at half maximum (FWHM) of the Ly-$\alpha$ line in the LAE sample of \citet{Kashikawa 2011} is $12.4 \pm 5.0$ {\AA}. Emission lines of variable flux were generated for the conditions of the observations, which allowed us to determine a detection limit\footnote{We fitted each simulated emission line with a Gaussian function to test whether a line was present, then by a straight line (test of the absence of a line). The detection limit is defined as the flux for which 80\% of the simulations have $\chi^2/ndf (Gaussian) = \chi^2/ndf (line)$.} of $\sim$10$^{-18}$ erg~cm$^{-2}$~s$^{-1}$. 

%Since no continuum is detected, we note that it is impossible to constrain the Equivalent Width (EW) of the Ly-$\alpha$ emission line.

%
%______________________________________________________________
%
\section{Discussion}
\label{sec:disc}

\subsection{Constraints on the properties of the GRB host galaxies at $z>5$}
\label{sec:const}

From the UV luminosity limits at $\sim$1500 {\AA}, we derived upper limits to the SFRs of the GRB hosts in our sample, using the \citet{Kennicutt 1998}, \citet{Moustakas 2006}, and \citet{Savaglio 2009} conversion relations. These various relations give results that are consistent to within a factor of about two. The resulting SFR values, reported in table~\ref{table:2}, do not exceed a few M$_\odot$~yr$^{-1}$.

Similarly, the null detection in the spectroscopic observation of the GRB 080913 host did also allow us to derive a constraint on the SFR of this object. At $z = 6.7$, the line flux limit corresponds to a Ly-$\alpha$ luminosity limit $L$(Ly-$\alpha$) $\lesssim$ 0.5$\times 10^{42}$ erg~s$^{-1}$. Using the empirical relation from \citet{Totani 2006}, $SFR$(Ly-$\alpha$) = $9.1 \times 10^{-43}$ L(Ly-$\alpha$)~M$_\odot$~yr$^{-1}$,  this corresponds to an extinction-uncorrected SFR of $\lesssim$ 0.5 M$_\odot$~yr$^{-1}$, in good agreement with the $<$ 0.9 M$_\odot$~yr$^{-1}$ upper limit derived from the photometric data (table~\ref{table:2}). Since the Ly-$\alpha$ line is usually strongly attenuated, even at high redshift, the photometric limit is probably more reliable than the spectroscopic one, and this is therefore the limit that we adopt.

For the sake of this discussion, we used of published results on another high-redshift GRB host, GRB 050904, at $z = 6.3$. This host could not be detected in deep HST and Spitzer observations \citep{Berger 2007}. Significant dust attenuation, $A_{1400 \AA} \approx 1.2$ magnitude, was first inferred from  the GRB afterglow observations of \citet{Berger 2007}, which agrees with the one deduced from the afterglow absorption spectrum \citep{Kawai 2006}. However, a reanalysis of the GRB 050904 afterglow observations questions the evidence of dust in this host. \citet{Zafar 2011} found a lack of even moderate attenuation, while \citet{Stratta 2011} again found some evidence of dust attenuation. %As explained in section  \ref{sec:lim-photo}, we assume that this dust extinction is representative of the dust extinction in the galaxy host as a whole. The UV continuum and SFR limits given in table~\ref{table:2} are corrected from this dust attenuation.
In the absence of a clear consensus, the most conservative limits given by the assumption of significant dust attenuation \citep{Berger 2007} are considered in this paper (table~\ref{table:2}). These do not affect our - essentially qualitative - conclusions.

\begin{figure}
 \resizebox{\hsize}{!}{\includegraphics{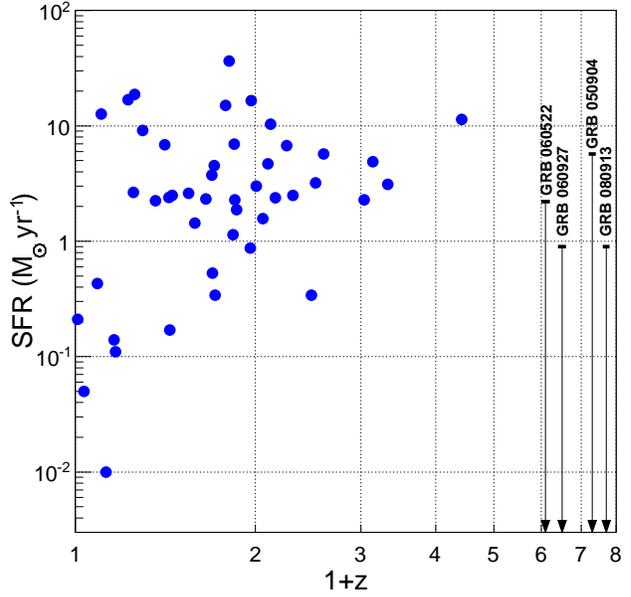}}
 \caption{Upper SFR limits (black arrows) for GRB 080913, GRB 060927, and GRB 060522 (these observations) and GRB 050904 \citep{Berger 2007}. The dots correspond to the SFR values of 46 lower redshift GRB hosts \citep{Savaglio 2009}.}
 \label{fig:sfr}
\end{figure}

The final sample used in this paper consists of four out of the five spectroscopically confirmed GRBs at $z>5$. Albeit based on low numbers, we assume that this sample is nonetheless representative of the GRB host population at high redshift, allowing us to discuss the properties of these objects. The constraints on the UV absolute magnitudes and derived SFR values of the four objects in this sample are reported in table~\ref{table:2}. These SFR limits are also shown in figure~\ref{fig:sfr} together with the SFR values of lower redshift GRB hosts. They are among the strongest constraints ever reported for high-redshift GRB host galaxies \citep{Tanvir 2012}.

With the SFR-mass relations given by \citet{Savaglio 2009} and \citet{Gonzalez 2011}, we infer an upper limit to the mass of the GRB hosts of $\sim10^{9.2}$ M$_\odot $ for a SFR lower than $\sim 1$ M$_\odot~yr^{-1}$ . 

\begin{table}
\caption{Limits on the UV absolute magnitude and SFR of four GRB host galaxies at high redshift, $z>5$. This sample represents two-thirds of the known GRB host galaxies in this redshift domain. }
\label{table:2}
\centering
\begin{tabular}{c c c c c }
\hline\hline
Host galaxy & Redshift & M$_{1500\AA}$ & SFR\\
\hline
GRB 080913 & 6.7 &  $>$ -19.4  &  $<$ 0.9 M$_\odot$~yr$^{-1}$\\
GRB 060927 & 5.5 &  $>$ -19.6  &  $<$ 0.9 M$_\odot$~yr$^{-1}$\\
GRB 060522 & 5.1 &  $>$ -20.5  &  $<$ 2.2 M$_\odot$~yr$^{-1}$\\
\hline
GRB 050904\tablefootmark{a} & 6.3 &  $>$ -20.7  &  $<$ 5.7 M$_\odot$~yr$^{-1}$\\
\hline
\end{tabular}

\tablefoottext{a}{From \citet{Berger 2007}, limits are corrected for dust attenuation.}
\end{table}

\subsection{Comparison with low redshift GRB host galaxies}

We now compare the properties of the GRB host galaxies at high redshift derived in the previous section to their properties at lower redshift. Adopting the \citet{Savaglio 2009} SFR data (figure~\ref{fig:sfr}), the probability that four observations give no detection is relatively low\footnote{We computed the cumulative distribution function, $CDF(SFR)$, of the SFR distribution at $z<1$ from \citet{Savaglio 2009}. This function gives the probability of observing a GRB in a galaxy with a $SFR$ below a given value. The probability that four observations give no detection is $\prod CDF(SFR)=0.80\times0.61\times0.36\times0.36\approx0.06$.}, $\sim$6 \%. This suggests that the GRB host galaxies have a lower SFR at redshifts $>5$ than they have at redshifts $\lesssim1$. This is not a surprise in itself as it is now well-known that the SFR decreases with redshift in the general galaxy population.

\subsection{Comparison with high-redshift galaxies}

We now turn to our comparison of the properties of the GRB host galaxies at $z > 5$ with the properties of the normal galaxy population at the same redshifts. To do this, we assumed that GRBs trace star formation in galaxies at $z > 5$ in an unbiased way, as at lower redshifts \citep{Jakobsson 2005}. Selecting galaxies based on the past occurrence of a GRB therefore introduces a bias towards the high luminosity ones, in proportion to their UV luminosities. Adopting the \citet{Jakobsson 2005} formulation, the luminosity function of GRB host galaxies is proportional to $ \Phi(z,M_{UV}) \times 10^{-0.4 \times M_{UV}}$, where $M_{UV}$ is the UV absolute magnitude and $\Phi(z,M_{UV})$ the luminosity function of the normal galaxy population. The probability $P(z,<M_{UV})$) that a GRB host galaxy at redshift $z$ is brighter than $M _{UV}$ is then
\begin{equation}
P(z,<M_{UV})= \frac{\int^{M_{UV}}_{-\infty}\Phi(z,M) \times 10^{-0.4 \times M} dM}{\int^{-10}_{-\infty} \Phi(z,M) \times 10^{-0.4 \times M} dM}$$
\end{equation}

, where $P(z,<M_{UV})$ is sensitive to two parameters, the upper limit to the integral in the normalization factor (denominator) and the LF faint-end slope, both parameters defining the contribution of the faint, but numerous, small galaxies. For the upper limit to the integral, we adopted M$_{UV} = -10$, as suggested by \citet{Bouwens 2011}, on the grounds that galaxies of such faint luminosities are probably suppressed due to the combined effects of UV background, SNe feedback, and inefficient gas cooling \citep{Read 2006,Dijkstra 2004}. For the LF itself, we use the LF from  \citet{Bouwens 2011} defined at a rest-frame wavelength of 1600 {\AA} and redshifts of about five, six, and seven. We note that despite the high-redshift LFs being relatively well-characterized, the remaining uncertainties, in particular in the faint-end slope, introduce errors in the results that do not however affect our conclusions. 

\begin{figure}
 \resizebox{\hsize}{!}{\includegraphics{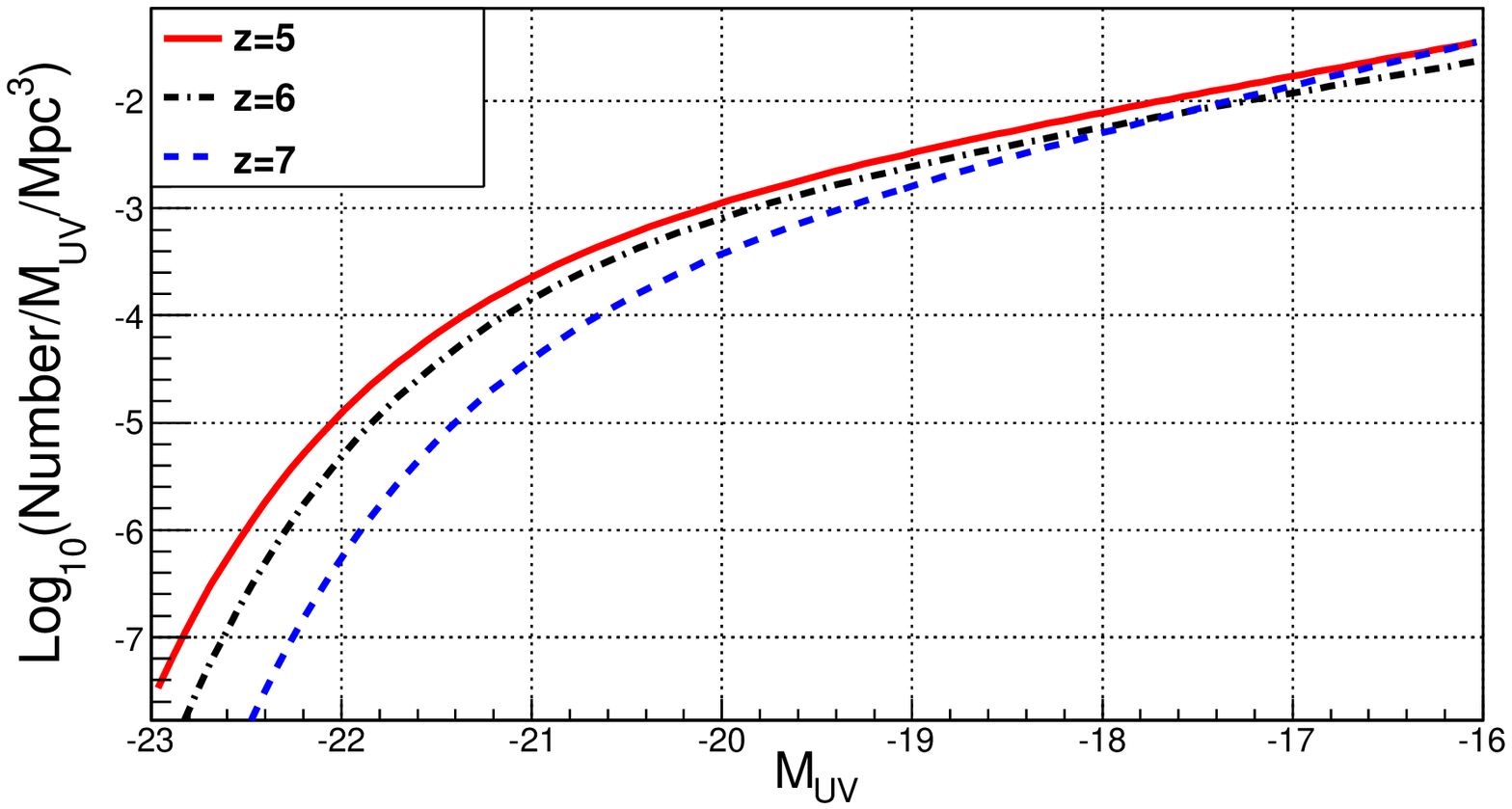}}
 \resizebox{\hsize}{!}{\includegraphics{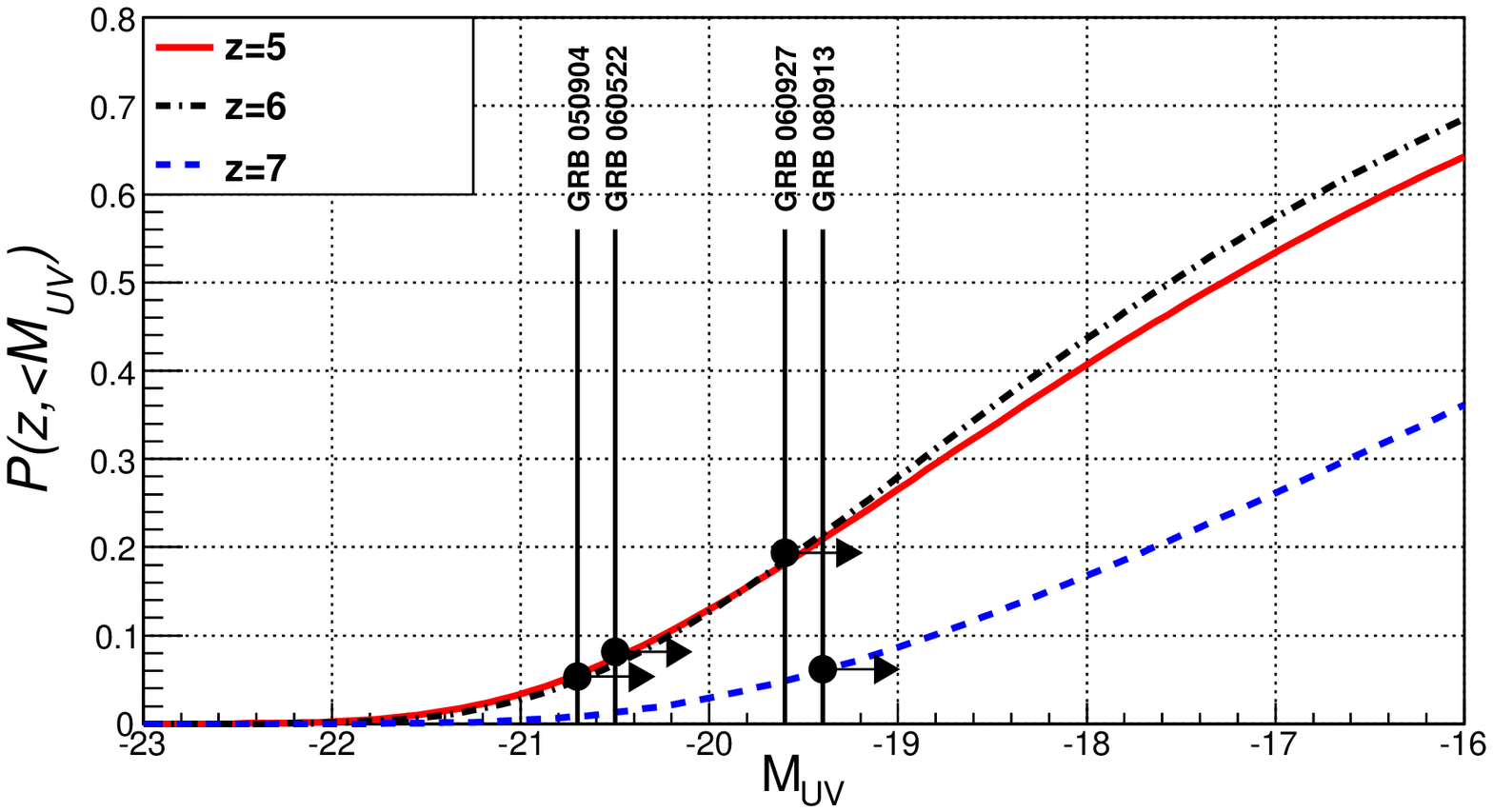}}
\caption{UV luminosity function of the high-redshift galaxies from \citet{Bouwens 2011} (top) and the probability, $P(z,<M_{UV})$, that a GRB host galaxy at redshift $z$ is brighter than $M _{UV}$ (bottom). For example, the probability that a GRB host galaxy at  $z=5.5$ is brighter than $M_{UV} = -19.6$, as for the GRB 060927 host (table \ref{table:2}), is $\sim$19 \%.}
 \label{fig:lf}
\end{figure}

With these assumptions, we computed $P(z,<M_{UV})$. Our results are shown in figure~\ref{fig:lf}, from which we were able to derive the probability that four observations give no detections. It is relatively high, 66\%\footnote{The probability that four observations give no detection is $\prod (1.0-P(z,<M_{UV}))=(1.0-0.19)\times(1.0-0.08)\times(1.0-0.06)\times(1.0-0.05)\approx0.66$.}, meaning that our null detections are consistent with the properties of the general galaxy population. We were unable to infer from our results whether the GRB host galaxies in our sample either differ from or are representative of the general galaxy population at similar redshifts.

Clearly, detecting galaxies at $z> 5$, and therefore GRB host galaxies, is a challenge that requires extremely deep observations. For instance, if the population of GRB host galaxies is representative of the population of normal galaxies, the probability of detecting a host galaxy counterpart  to a GRB at $z \sim 5.5$ and brighter than M$_{UV}\approx -17.5$ is 50\%. This corresponds to a detection limit of about two magnitudes deeper than the limits reported in this work.

\subsection{Constraining the nature of the GRB 080913 host galaxy}

We then considered GRB 080913, following a similar approach to the above to evaluate the statistical significance of the non-detection of the Ly-$\alpha$ line on its spectrum, and the conclusions that can be tentatively drawn from this - admittedly single - non detection.

The upper limit to the Ly-$\alpha$ luminosity of GRB 080913 host at $z=6.7$ is $\lesssim 0.5 \times 10^{42}$ erg~s$^{-1}$ (see section \ref{sec:lim-spectro}). This can be compared to the luminosity of the Ly-$\alpha$ line in other high-redshift objects. The GRB 090205 host galaxy at $z=4.6$, observed by \citet{D'Avanzo 2010}, has for example a Ly-$\alpha$ luminosity equal to $4.3\times10^{42}$ erg~s$^{-1}$. This value is in the range of the Ly-$\alpha$ luminosities ($1-15 \times 10^{42}$ erg~s$^{-1}$) observed in lower redshift ($1.9<z<3.4$) GRB hosts  exhibiting Ly-$\alpha$ emission (table~\ref{table:3}). The more general population of $z\approx 6.6$ LAEs exhibits similar Ly-$\alpha$ luminosities ($2 -39 \times 10^{42}$ erg~s$^{-1}$) (see e.g. \citet{Ouchi 2010} and \citet{Kashikawa 2011}).

\begin{figure}
 \resizebox{\hsize}{!}{\includegraphics{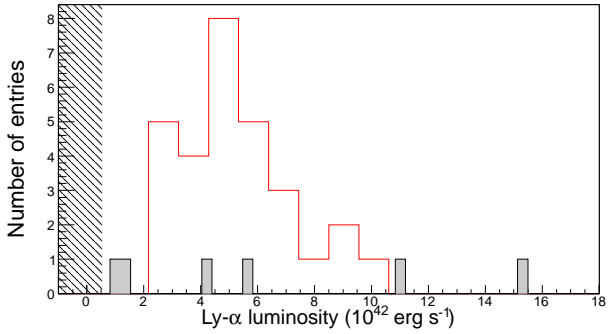}}
\caption{Distribution of Ly-$\alpha$ luminosities in $z = 6.6$ LAEs \citep{Kashikawa 2011} (red line). The Ly-$\alpha$ luminosities of the GRB hosts listed in table~\ref{table:3} are represented by the filled gray rectangles. Our limit to the Ly-$\alpha$ luminosity of the GRB 080913 host is indicated by the vertical dashed line.}
\label{fig:lyalpha}
\end{figure}

To compare the Ly-$\alpha$ luminosity limit of GRB 080913 to the general distribution of Ly-$\alpha$ luminosities in high-redshift objects, we considered the LAE population at $z\approx 6.6$ from \citet{Kashikawa 2011}. The distribution of Ly-$\alpha$ luminosities in this population is represented in figure~\ref{fig:lyalpha}. Taking this distribution as representative of the LAE population at the GRB 080913 redshift, we derived a probability\footnote{We fitted the distribution of Ly-$\alpha$ luminosities at $z\approx 6.6$ from \citet{Kashikawa 2011} with a normal distribution and computed the probability of having a luminosity fainter than $0.5 \times 10^{42}$ erg~s$^{-1}$.} of not detecting the host galaxy of GRB 080913, if it was a  Ly-$\alpha$ emitter, of $\sim 4 \%$.

We infer from this simple analysis that the host galaxy of GRB 080913 is probably not a Ly-$\alpha$ emitter. This is consistent with the observations from \citet{Jakobsson 2011}, indicating that the fraction of Ly-$\alpha$ emitters among GRB host galaxies at redshifts $1.8 < z < 4.5$ is only $\sim 37 \%$.

%We infer from this simple analysis that the host galaxy of GRB 080913 is probably not a Ly-$\alpha$ emitter. This is somewhat in agreement with the trend of decreasing Ly-$\alpha$ emission observed at high-redshifts ($> 6.5$), possibly due to an increasing neutral fraction of the IGM. Evidences of this tentative trend come from the observation of high-redshift LAEs (see e.g. \citet{Clement 2011}), and of the fraction of Ly-$\alpha$ emitters among LBGs \citep{Pentericci 2011}. As a consequence, searching for Ly-$\alpha$ emission in the spectra of high-redshift GRB hosts may not be the best strategy to detect these hosts.

\begin{table}
\caption{Ly-$\alpha$ luminosities of GRB host galaxies.}
\label{table:3}
\centering
\begin{tabular}{c c l l }
\hline\hline
GRB	& Redshift &	Ly-$\alpha$ luminosity	 &	Reference \\
		&			& (10$^{42}$ erg~s$^{-1}$) & \\
\hline
080913 & 6.69&      $<0.5$  & This work \\
\hline
971214 & 3.42&		$70.0\pm7.9$		&\citet{Kulkarni 1998}\\
000926 & 2.04&		$5.8\pm0.4$	&\citet{Fynbo 2002}\\
011211 & 2.14&		$1.0\pm0.3$		&\citet{Fynbo 2003}\\
021004 & 2.33&		$11.1 \pm 2.6$	&\citet{Jakobsson 2005}\\
030226 & 1.99&		$<72.3$			     &\citet{Jakobsson 2005}\\
030323 & 3.37&		$1.3 \pm 0.1$		&\citet{Vreeswijk 2004}\\
030429 & 2.66&		$15.3 \pm 7.9$	&\citet{Jakobsson 2004}\\
050904 & 6.29&	     $< 0.8$			&\citet{Totani 2006}\\
090205 & 4.65&		4.3			          &\citet{D'Avanzo 2010}\\
\hline
\end{tabular}
\end{table}

%
%______________________________________________________________
%
\section{Summary}

We have performed deep near-infrared photometric observations targeting the rest-frame UV continuum (\hbox{$\sim$1500 {\AA}}) in three GRB host galaxies at redshifts $>5$. In addition, for one of these objects (the host of GRB 080913), we have performed deep spectroscopic observations targeting the Ly-$\alpha$ line. Making use of results available in the literature for a fourth object, we have compiled a sample of four distant GRB host galaxies, from the five most distant GRBs at $z>5$ for which spectroscopic redshifts are available.

None of the GRB host galaxies are detected down to the sensitivity limit of our observations. From the UV rest-frame luminosity limits derived from these null detections, we have inferred that the four objects in the sample have SFRs that are statistically lower than at $z\lesssim1$. Assuming that our sample is representative of the population of GRB host galaxies at $z \gtrsim 5$, we therefore conclude that GRB host galaxies at $z\lesssim1$ and at $z \gtrsim 5$ have statistically different SFRs. Conversely, these null detections are statistically compatible with the properties of the general galaxy population at $z \gtrsim 5$. Deeper observations would be required to investigate whether the two populations have different properties.

We did not detect Ly-$\alpha$ emission in the host of GRB 080913 at $z = 6.7$, and from the detection limit reached by our observations we infer that this object is probably not a Ly-$\alpha$ emitter.

These results indicate that the observations of GRB host galaxies at $z > 5$ is a challenging endeavor, and systematic observations of this population of galaxies will need to await the next generation of large telescopes (JWST, ELTs). Meanwhile, real-time photometric and spectroscopic observations of GRB afterglows will remain a powerful tool to characterize the farthest collapsed objects in the Universe and probe their environments.

\begin{acknowledgements}
We thank the anonymous referee for helpful comments. We are grateful to Chris Lidman for very useful suggestions, and all the GROND team at MPE for providing the finding chart of GRB 080913. We also express our thanks to the Paranal Observatory staff for their excellent support and assistance. This research is based on observations made with ESO Telescopes at the Paranal Observatory under programs 60.A-9402(A), 085.A-0418(A), and 085.A-0418(B). It has also made use of the GHostS database (www.grbhosts.org), which is partly funded by Spitzer/NASA grant RSA Agreement No. 1287913. .
\end{acknowledgements}

\bibliographystyle{aa} % style aa.bst

\end{document}